\documentclass[times, twoside, watermark]{zHenriquesLab-StyleBioRxiv}

\usepackage{hyperref}
\usepackage{color}
\usepackage{tabularray}
\usepackage[breakable, skins]{tcolorbox}
\usepackage{enumitem}
\leadauthor{F-BIAS team} 

\usepackage{soul} 
\definecolor{lightyellow}{rgb}{1.0, 1.0, 0.8} 
\sethlcolor{lightyellow}

\begin{document}

\title{F-BIAS: Towards a distributed national core facility for Bioimage Analysis.}
\shorttitle{A national core facility for Bioimage Analysis}



\author[1]{Mélodie Ambroset}
\author[2]{Marie Anselmet}
\author[3]{Clément Benedetti}
\author[4]{Arthur Meslin}
\author[5]{Aurélien Maillot}
\author[6]{Christian Rouvière}
\author[7]{Guillaume Maucort}
\author[8]{Marine Breuilly}
\author[2]{Marvin Albert}
\author[9]{Gaëlle Letort}
\author[10]{Antonio Trullo}
\author[3]{Volker Bäcker}
\author[11]{Jérôme Mutterer}
\author[7]{Fabrice Cordelières}
\author[12]{Thierry Pécot}
\author[2]{Minh-Son Phan}
\author[2]{Stéphane Rigaud}
\author[1]{Magali Feyeux}
\author[1]{Perrine Paul-Gilloteaux}
\author[4 \Letter]{Anne-Sophie Macé}
\author[2 \Letter]{Jean-Yves Tinevez}

\affil[1]{Nantes Université, CHU Nantes, CNRS, INSERM, BioCore, US16, SFR Bonamy, 44000 Nantes, France} 
\affil[2]{Institut Pasteur, Université Paris Cité, Image Analysis Hub (IAH), 75015 Paris, France} 
\affil[3]{Montpellier Ressources Imagerie, BioCampus, University of Montpellier, CNRS, INSERM, 34090 Montpellier, France} 
\affil[4]{Cell and Tissue Imaging Facility (PICT-IBiSA), Institut Curie, PSL Research University, CNRS, 75005 Paris, France} 
\affil[5]{Institut Curie, Laboratory of Genetics and Developmental Biology, PSL Research University, INSERM U934, CNRS UMR3215, 75248 Paris, France} 
\affil[6]{CBI Image Processing, Centre de Biologie Integrative, Université de Toulouse, CNRS, 31062 Toulouse, France} 
\affil[7]{University of Bordeaux, CNRS, INSERM, France-BioImaging INBS, Bordeaux Imaging Center (BIC), UMS 3420, US 24, 33076 Bordeaux, France} 
\affil[8]{CIQLE, LyMIC, LABEX CORTEX, Université Claude Bernard Lyon 1, Structure Fédérative de Recherche santé Lyon-Est CNRS UAR3453/Inserm US7, 69008 Lyon, France} 
\affil[9]{Department of Developmental and Stem Cell Biology, Institut Pasteur, Université de Paris Cité, CNRS UMR 3738, 75015 Paris, France} 
\affil[10]{Institut de Génétique Moléculaire de Montpellier, University of Montpellier, CNRS-UMR 5535, 34293 Montpellier, France} 
\affil[11]{Institut de biologie moléculaire des plantes, CNRS, Université de Strasbourg, 67000 Strasbourg, France} 
\affil[12]{Facility for Artificial Intelligence and Image Analysis (FAIIA), Biosit UAR 3480 CNRS-US18 INSERM, Rennes University, 35042 Rennes, France} 
\affil[*]{Authors are listed in reverse order of their joining the F-BIAS project.}

\maketitle

\definecolor{verylightgray}{RGB}{250, 250, 250}
\begin{tcolorbox}[boxrule=0.5pt, colback=verylightgray]
    \footnotesize
    \noindent \textbf{Key points}
    \begin{itemize}[wide = 0pt]
        \itemsep0em 
        \item We describe our experience in creating a nation\-wide core facility offering services in bioimage analysis.
        \item This virtual facility federates existing resources scattered across the territory and within different research institutions.
        \item It attracted and retained analysts thanks to the scientific added value it brought to them. 
        \item The distributed core facility efficiently addresses the needs of imaging-based research projects, in particular for researchers previously without access to image analysis expertise.
        \item It also mitigates the risk of isolation for analyst members that are often the sole expert in image analysis in their local teams.
        \item We identify and share the critical components necessary for the success of similar endeavors, to facilitate reproducibility.
    \end{itemize}    
\end{tcolorbox}

\begin{abstract}
\noindent We discuss in this article the creation and organization of a national
core facility for bioimage analysis, based on a distributed team. F-BIAS
federates bioimage analysts across France and relies on them to deliver services 
to the researchers of this territory. The main challenge in implementing this 
structure is to ensure significant scientific value to the analysts, thereby 
encouraging their active participation and persuading their respective host teams 
to support their involvement. F-BIAS accomplished this by creating a professional 
network that mitigates the negative effects of isolation experienced by its 
members, who are often the sole bioimage analyst within their local teams, and 
fosters the development of their technical skills. In a second phase we 
capitalized on F-BIAS to create a virtual, remotely-operating core facility for 
bioimage analysis, offering consultations and collaborative project services to 
the scientific community of France. The services are organized so that they also 
contribute to the technical proficiency of the analysts. To promote the creation 
of similar structures, we present and discuss here the organization of this 
nationally distributed bioimage analysis service core, highlighting successes and 
challenges.
\end {abstract}

\begin{keywords}
\noindent bioimage analysis | national core facility | distributed team | technical skills development
\end{keywords}

\begin{corrauthor}
\texttt{tinevez{@}pasteur.fr, anne-sophie.mace{@}curie.fr}
\end{corrauthor}

\begin{acronyms}
CNRS: Centre national de la recherche scientifique.
GloBIAS: Global BioImage Analysts' Society.
INSERM: Institut national de la santé et de la recherche médicale.
NEUBIAS: Network of European BioImage Analysts.
RMS: Royal Microscopical Society. 
RTmfm: Réseau Technologique Microscopie Fluorescence Multidimensionnelle.
\end{acronyms}

\section*{Introduction}

\subsection*{The growing need for support in bioimage analysis}
\label{the-growing-need-for-support-in-bioimage-analysis}

Tremendous scientific progress has been made in recent years to allow the exploration of biological systems from the wet to the dry parts of the investigation. 
Image analysis applied to biology - bioimage analysis - has become a central component of imaging-based research projects~\cite{miura_survey_2021, senft_biologists_2023}. 
It involves a large range of methodological approaches, from image processing (denoising, registration, deconvolution...) through visualization to quantification (segmentation, features measurement, tracking…), optionally up to statistical and mathematical analysis~\cite{eliceiri_biological_2012, meijering_imagining_2016}. 
A major task for the bioimage analysis community focus is to create and deploy scientific tools to enable biologists to analyze their own data without requiring advanced computing expertise~\cite{cimini_community_2023}. 
Nonetheless, the increasing number of available scientific software tools~\cite{haase_hitchhikers_2022}, the need for more and more powerful computing resources to analyze large datasets~\cite{Wolff2018}, the use of deep-learning solutions and the fast evolution of the artificial intelligence field~\cite{weigert_content-aware_2018, belthangady_applications_2019, moen_deep_2019} makes it very difficult for non-experts to navigate through and get familiar with these techniques~\cite{cimini_community_2023}. 
A poll conducted in 2015~\cite{miura_survey_2021} showed that 59.3\% of respondents (in majority life scientists or microscopists) considered image analysis to be the most difficult task in their work, which was reaffirmed recently~\cite{sivagurunathan_bridging_nodate}. 
It also revealed that 39\% of respondents lacked access to any support in image analysis. 
The gap between life scientists that have the knowledge of the biological data and analysts that understand the computational techniques remains a challenge~\cite{sivagurunathan_bridging_nodate, lambert_towards_2023, chen_when_2023} and is widening in the artificial intelligence era~\cite{cimini_community_2023, laine_avoiding_2021}. 

Various initiatives, such as NEUBIAS~\cite{martins_highlights_2021}, have played a crucial role in shaping a new position in academic research: the bioimage analyst. Analysts support research projects by creating customized analysis pipelines, either by using existing tools and techniques or by adapting and developing new algorithms to address the specific biological question. They aim at training and advising biologists, facilitating access to computing resources and providing technical solutions through the development of new tools or the fine-tuning of existing solutions to specific problems~\cite{soltwedel_challenges_2024}. 
They act as experts or as an interface with experts, that develop, adapt or assess the values of novel methods, focusing on the specific needs of a life science project~\cite{lambert_towards_2023}.

\subsection*{The isolated analyst}
\label{the-isolated-analyst}

Core facilities are today a well-established part of the research landscape in academia. 
They are run by dedicated staff and provide expertise and assistance to research projects in biology~\cite{lewitter_establishing_2009}. 
While centralizing computational resources for bioinformatics is well established~\cite{lewitter_need_2009}, bioimage analysis core facilities - dedicated structures that house analysts and provide image analysis services to a broad scientific community~\cite{soltwedel_challenges_2024, deschamps_better_2023} - are only emerging.
Most of the time analysts, when they are present, are embedded in large microscopy cores. 
This has some clear advantages in a holistic approach: bioimage analysis is indissociable from imaging. 
For instance, a challenging bioimage analysis project can be made simpler by choosing the right imaging modality. 

Bioimage analysis has been blooming over the last decade.
A large number of important new tools and algorithms were frequently published, making it impossible for a single analyst to master or even know of them all. 
Bioimage analysts often find themselves isolated and have nobody to consult or to exchange ideas with. 
However, a professional environment comprising peers is crucial to progress technically, to improve their skills and to stay competitive against the rapidly changing technological landscape~\cite{renaud_staying_2024}. 
It could also be beneficial when working on user projects to ask for help and feedback. 
Without such a stimulating environment, analysts might look for better prospects outside of academia~\cite{deschamps_better_2023}. 
This situation is very similar to what bioinformaticians endured a decade ago~\cite{watson_guide_2013}.
To face these problems, dedicated research infrastructures were built ~\cite{lewitter_establishing_2009, 
lewitter_need_2009, dillies_pasteur_nodate} or large professional societies 
dedicated to bioinformatics~\cite{noauthor_sfbi_nodate, noauthor_iscb_nodate} were created.

Several initiatives aim at mitigating the adverse effect of the isolation, such as the creation of the Bioimage Analysis forum~\cite{rueden_scientific_2019} or image analysis consortiums like NEUBIAS~\cite{martins_highlights_2021} or GloBIAS~\cite{noauthor_globias_nodate}, and national networking initiatives, usually as subgroups of national microscopy networks, such as RMS (UK) or RTmfm (France). 
However, they do not substitute for a team environment where informal and frequent discussions would take place.
Also, their scope is much broader than the provision of bioimage analysis services, a topic that is still taking off. 
To overcome analyst isolation, smaller networks with a more focused scope and frequent, tailored interactions are necessary.

\subsection*{The F-BIAS initiatives}
\label{the-f-bias-initiatives}

\begin{figure}[!t]
    \centering
    \includegraphics[width=0.4\textwidth]{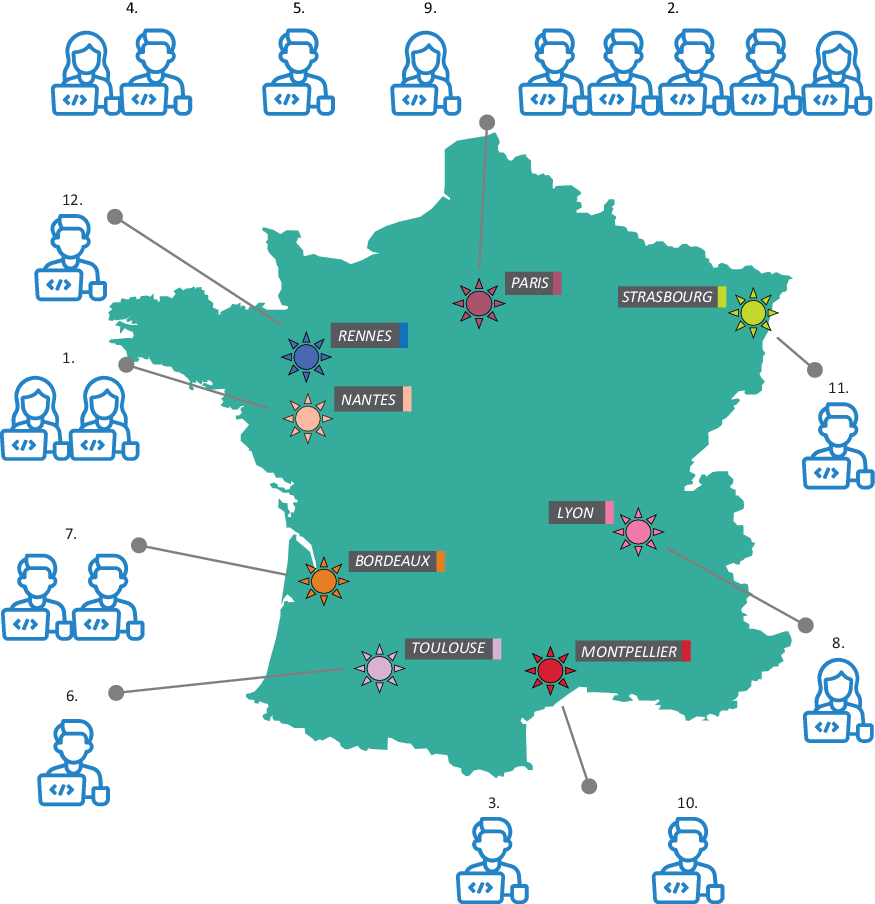}
    \caption{The geographical repartition of F-BIAS members and their host structure (here in 2024). 
    F-BIAS federates willing FBI research engineers whose main mission is to provide bioimage analysis services to the scientific community. 
    With few exceptions, the main model is for bioimage analysis services to be provided by 1-2 analysts embedded within a larger structure dedicated to imaging.
    Numbers above the analysts' icons correspond to their respective affiliations listed in the preamble of this article.}
    \label{fig:FigureN1}
\end{figure}

France-Bio\-Imaging (FBI), the French node of Euro\-Bio\-Imaging~\cite{pfander_euro-bioimaging_2022}, is the French nationwide infrastructure for biological imaging~\cite{fbi_about_page}. 
It is a distributed infrastructure that provides its users with quick access to cutting-edge and innovative microscopy, associated expertise, labeling methods, image analysis, and tailored training. 
It is composed of microscopy core facilities but also of research and development teams in these different fields in order to foster technological transfer of innovation to users.
Many of the analysts working in French core facilities followed the isolated analyst model described above (Figure~\ref{fig:FigureN1}). 
This led us to build a distributed structure, supported by FBI, to mitigate the issues of isolation on the scale of a country. In turn, it opened up an opportunity to build a nationwide, virtual, remotely-operating core facility for bioimage analysis. 
In the following we report the creation of F-BIAS: \textit{France-BioImaging Analysts} the federation of analysts working in French core facilities. 
F-BIAS was initially created to have a strong scientific value to analysts. 
In a second phase, it was used to offer bioimage analysis services to the French scientific community as a distributed core. 
We describe below how F-BIAS was designed, what rules were used and what can be achieved with such an organization. We also discuss its limitations and challenges and share our experience, hoping that it will be useful for other bioimage analysis communities over the world.

\section*{The deployment of a distributed team of analysts and the services it provides at a national level}
\label{the-deployment}

While developing F-BIAS, we encountered several challenges throughout the different phases of deployment, as outlined in Table~\ref{table-steps-challenges} and detailed below.

\definecolor{LunarGreen}{rgb}{0.215,0.254,0.223}
\definecolor{Black}{rgb}{0,0,0}
\definecolor{Nandor}{rgb}{0.286,0.337,0.298}
\definecolor{Tundora}{rgb}{0.262,0.262,0.262}
\definecolor{AthensGray}{rgb}{0.972,0.976,0.98}
\begin{table*}
    \centering
    \footnotesize
    \begin{tblr}{
      width = \linewidth,
      colspec = {Q[25]Q[50]Q[115]},
      cells = {m, fg=Tundora},
      row{odd} = {AthensGray},
      row{1} = {Nandor,fg=white},
      column{1} = {l},
      column{2} = {l},
      column{3} = {l},
      hlines,
      vlines = {Black},
      vline{1} = {-}{LunarGreen},
    }
        Steps &	Challenges & Approach \\
            
        \SetCell[r=3]{c}{Inception of the F-BIAS network } 
        & Defining a target community & Restricted to research engineers working in a FBI core facility, providing services to the scientific community and whose main occupation is bioimage analysis. \\  
        & Defining requirements for membership & Regular attendance at monthly meetings and open desks is required. \\  
        & Recruiting initial members & Personal 1-1 calls with tentative members, advocating the network and collecting expectations, focused on the needs of isolated analysts. \\  

        \SetCell[r=3]{c}{Monthly meetings organization} 
        & Planning the meeting sessions & Poll at the beginning of the year to define the optimal meeting time, and sending invitations. \\  
        & Establishing rules for exchange of information & Schedule private meetings (without users) and secure a commitment from attendees to disclose any conflict of interest related to projects presented during the meeting. \\ 
        & Encouraging attendance & Facilitate continuous discussions on the instant chat system between sessions, and send a reminder one week before each meeting. Restrict meetings content to analysts' occupation: get advice and help on user projects,  discuss new tools or techniques developed by a member or published by others. \\  

        \SetCell[r=2]{c}{Yearly meetings organization} 
        & Coordinating logistics & Hosted and organized by an analyst member in the core facility they belong to. \\ 
	& Defining the meeting scientific content & Use the instant chat system for discussions to define the program, which will focus on technical talks for upskilling and general discussions on organizing the bioimage analysis core facility's work. \\  
    
        \SetCell[r=5]{c}{Open desk sessions (Figure~\ref{fig:FigureN2}A)} 
        & Planning the next session & Fixed during  monthly meetings to have as many available analysts as possible. One open-desk session can last 1-3 hours, have 1-6 slots of one hour, with possibly 2 slots running in parallel. \\ 
        & Communicating about the session & Announcement to all FBI users by the FBI External Affairs Manager \& creation of an online page on FBI website for enrollment. \\  
        & Assembling the analysts group for one slot & Analysts are mixed as much as possible based on their profiles to achieve the following: a)~encourage them to learn and explore different approaches and tools, b)~provide users with a broader range of ideas and perspectives.\\ 
        & Maximizing the efficiency and the scientific value of the open desk slot  & Users are requested to send example data in advance so that analysts can work on them. \\ 
        & Keeping track of the session content &	Shared Google Doc file with user information as well as the proposed solution, reasonably detailed. \\  
        
        \SetCell[r=7]{c}{Collaborative projects (Figure~\ref{fig:FigureN2}B)} 
        & Obtaining dedicated human resources & Advocate for the need for these resources within the FBI umbrella by showcasing the success of a proof-of-concept collaborative project.  \\
        & Defining a recharge policy & Discussions with the FBI executive board to define an hourly fee that can be subsidized by FBI. Discussions with members core facilities to implement the subsidized fee to F-BIAS collaborative projects. \\
        & Defining an entry point for user requests & Discussions with a user during an open desk session or via direct contact. The project request is then formalized by the user through the Euro-BioImaging Access Portal. \\
        & Choosing the analysts for collaboration & Discussion during monthly meeting to a)~check availability and expertise of analysts, b)~get some advices on the project and c)~potentially elect a mentor analyst that will support the collaborating analyst. \\  
        & Performing follow-up for the collaborative project & Regular meetings are held between the analyst and the user, the collaborating analyst and the mentor analyst, or all parties together. The collaborative analyst can also rely on the F-BIAS group. Project management will use the tools and policies of the analyst's core facility. \\
        & Sharing project data & Perspective: use of an OMERO database instance deployed with the help of FBI.Data project across the different member structures. \\  
        & Charging the user for the project &	Billing is managed by the host core facility of the collaborating analyst, based on the agreed-upon fees.  \\
    \end{tblr}
    \caption{List of the milestones, their challenges, and the approaches we used to address them.}
    \label{table-steps-challenges}
\end{table*}

\subsection*{Federating analysts in remote locations and fostering the development of their professional and technical skills}
\label{federating-analysts}

In France, research support positions are typically held by technicians or research engineers, who have a dedicated career path distinct from that of researchers. 
In core facilities, these positions can be either fixed-term contracts or permanent, and they are often funded by the institutions that host the core facilities (such as universities, Institut Pasteur, Institut Curie, etc.) or by research organizations (like CNRS, INSERM, etc.) that fund these cores. 
The mission of analysts in these positions is to support their local scientific community through the activities of the core facility in which they are embedded.

F-BIAS was created to federate analysts scattered across France in order to provide an adequate professional environment and to address the needs listed in the previous section.
To this aim, a few rules were defined. First, members must be bioimage analysts working in a core facility in France whose main mission is to deliver bioimage analysis services to the wide scientific community. 
Image processing scientists working in research labs cannot join F-BIAS as support is not part of their mission and their work mainly focuses on their own projects. 
Second, members have to contribute to the services of the nationwide core facility described below, mainly by participating in the open desk sessions. 
Third, meetings and exchanges are private and confidential. 
Analysts are asked not to disclose any information discussed during meetings unless agreed upon and to immediately voice concerns if they see a possible conflict of interest when a topic is brought up by a peer. 
This helps to create a safe discussion place where confidentiality issues are addressed on a case-by-case basis.

In practice, F-BIAS was initiated mid 2021 by inviting analysts working in FBI core facilities to join under the rules and scope outlined above.
As candidate members already held research support positions, no initial funding was required.
The initial effort involved convincing the analysts and their managers to allocate some of the analysts' time for this network.
Since some of the research organizations in France are national entities (such as CNRS and INSERM), they create incentives to participate in national initiatives like F-BIAS.
As such, F-BIAS started as virtual network, federating existing resources. 
The resulting network started with 10 members and has since expanded to include 20 members today (Figure~\ref{fig:FigureN1}). 
This turns out to be a manageable size, handy to maintain focused, lively, and efficient exchanges. 
All members have the same line of work with a common commitment to support a scientific community with bioimage analysis, but they show a  diversity of backgrounds (applied mathematics, biology, computer science, physics), domains of applications (cell biology, developmental biology, histopathology, microbiology, plant biology, …) and career stage. 
The initial activities involved establishing monthly meetings and deploying team chat tools to favor instantaneous exchange and structured communication. 
Meetings are prepared in advance to include 2-5 topics including 1/~sharing preview versions of tools and pipelines developed by members and collecting feedback, 2/~requesting help by peers on a user project, 3/~discussing an article or a new bioimage analysis tool. 
Approximately 30 meetings have taken place since the creation of F-BIAS.
Additional yearly in-person of 2-3 days and other gatherings during FBI meetings help to maintain team spirit. Most of the analysts are also active in other networks such as GloBIAS or CZI.
When a member wishes to learn a new technique, we inquire within the group to see if someone is knowledgeable in that area and willing to conduct a short demonstration or workshop at our annual or monthly meetings.
F-BIAS governance is horizontal and relies on two to three members for event organization, service coordination and reporting to FBI, our umbrella infrastructure.

\subsection*{From technical networking to bioimage analysis service provision}
\label{offering-consultations-the-fbi-remote-open-desks-on-bioimage-analysis}

F-BIAS aims to provide bioimage analysis services to the entire scientific community in France, particularly life scientists with no analyst onsite. 
To fulfill this mission, F-BIAS offers \textit{bioimage analysis consultations} to researchers in France in the form of an open desk, organized every other month, and communicated through FBI newsletter and website. 
For each open desk, two to six online sessions of 1 hour are planned, with users booking one of the sessions in advance. 
Users are required to upload images before the session with a description of their needs. 
Descriptions are then used to select 2 to 3 F-BIAS members that will join the user in an online session of 1 hour (Figure~\ref{fig:FigureN2}A). 
Analysts can offer advice on which tool to use, how to use it with online demonstration on the user data, implement a preliminary version of an analysis pipeline, or create small macros for analysis or task automation. 
Within the allocated time of the session, the goal is to address the user’s specific question and analysts are required to prototype the solutions and advice they provide. 
Most of the time, the different analysts propose different approaches. 
Consultations are free and users are expected to acknowledge the FBI infrastructure in their publications.

These consultations only require a few hours from the analysts, ensuring their participation does not compromise their daily responsibilities. 
This service started mid 2022 and the feedback from users has been overwhelmingly positive from the outset. 
Consultations are indeed highly beneficial to users as they provide the initial leverage needed for an image analysis project. 
They offer help to scientists without access to analysts, who critically need it. 
A collateral advantage was the unexpected positive impact on analysts’ expertise. 
As stated above, given the proliferation of bioimage analysis tools, it is impossible for a single analyst to be proficient in all of them. 
Each analyst in a session prototypes a different solution to the user’s problem. 
This allows the other analysts to learn about the advantages and limitations of the tools being discussed. 

A shared document is completed at the end of each session, summarizing the attending analysts and the solution proposed to the user as well as any raised issues (such as the need for better resolution or advices for improved imaging etc). 
This ensures the continuity if the user attends another open desk with different analysts.

\subsection*{Beyond consultations: Bioimage analysis collaborative projects services}
\label{beyond-consultations-bioimage-analysis-collaborative-projects}

Consultations are best suited when the image analysis task can be solved with existing tools or well-established pipelines. 
However, imaging-based research feeds on innovation. 
In some cases, researchers have questions that have never been addressed before and for which no tools are suited. 
A significant part of the analysts’ scientific added value is their ability to adapt existing tools and to create bespoke image analysis pipelines, in the framework of \textit{collaborative projects}. 
In the scope of bioimage analysis, this term relates to the elaboration and delivery of an image analysis pipeline or of image analysis results by the analyst to the user, with both parties closely working together. 
The project corresponds to an articulated request from the user but involves an original contribution and some scientific responsibility from the analyst. 
F-BIAS aims at offering such services at a national level.

While consultations require moderate time commitments, collaborative projects require a significant amount of the analyst time. 
Understandably, both analysts and their supervisors expressed caution when joining the initiative. 
The main mission of an analyst is to support the users of their local core facility.
We observed that all the initial members of F-BIAS were already fully occupied with local projects, which naturally limited their ability to accept new projects from outside their hosting core facilities. 
Consequently, we turned to FBI, acting as the national infrastructure umbrella, to provide dedicated resources.
An important factor in advocating for these resources was a proof-of-concept collaborative project established in 2023, involving an analyst, a user and a mentoring analyst in three distant cities.
It demonstrated the feasibility of delivering high-quality services remotely and integrating them into the existing service offers of host core facilities while sharing the work load and allowing mentoring between analysts. 
This success facilitated the allocation of two junior engineer positions dedicated to F-BIAS projects, financially supported by the FBI umbrella infrastructure, and that were embedded in two cores of F-BIAS.
Thanks to these recruitments, within the first six months of 2024, we were able to accept and plan eight requests for collaborative projects. 
Out of these eight requests, three were completed and one was deemed unfeasible. 
User requests cover a wide range of domains in biology and image analysis (Table~\ref{table-projects}), which highlights the diversity of expertise F-BIAS can offer, with a distributed team coming from diverse backgrounds.

\begin{figure}[!t]
    \centering
    \includegraphics[width=0.4\textwidth]{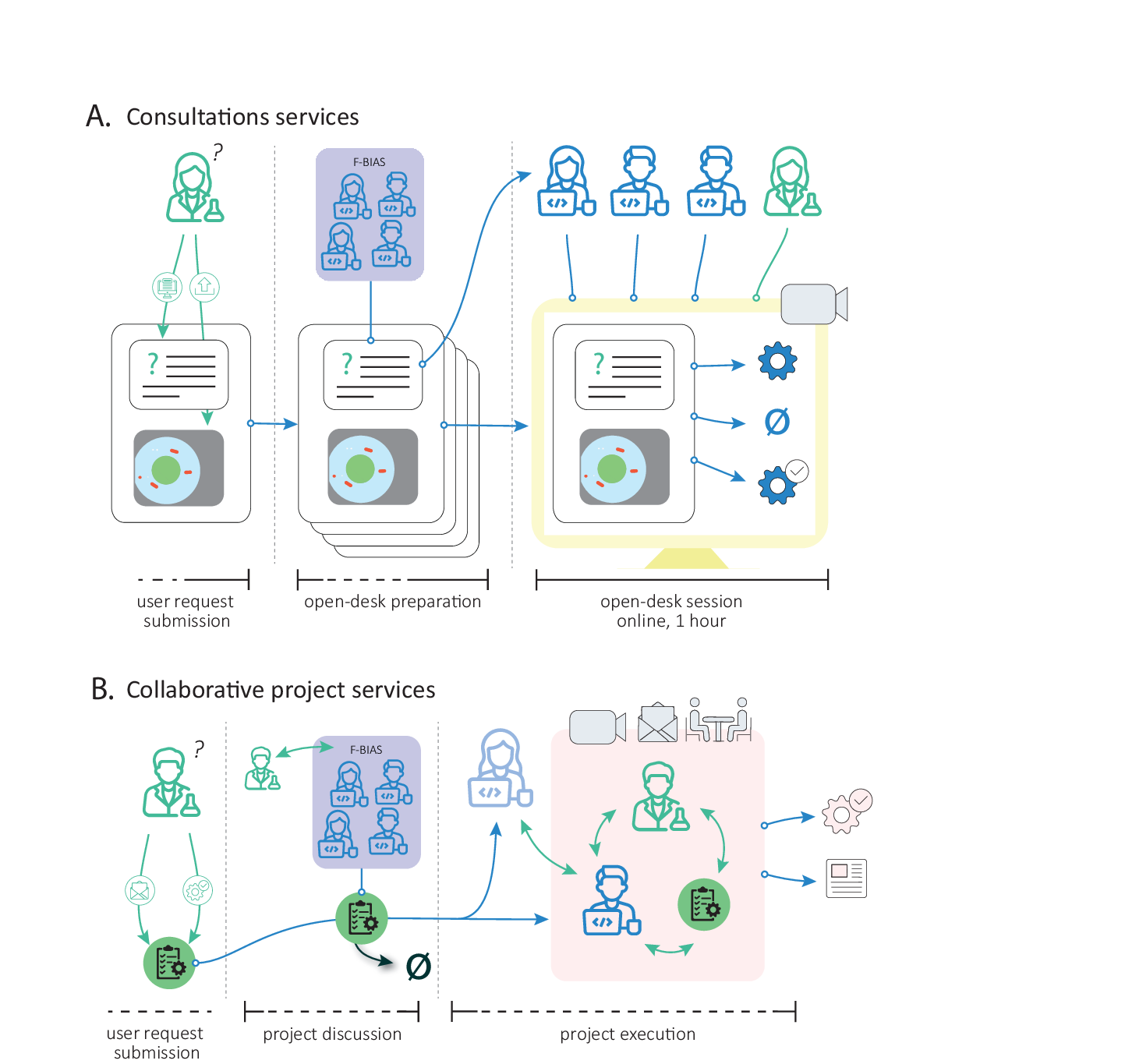}
    \caption{Overview of the bioimage analysis services provided by F-BIAS. \textbf{A.}~The consultations services take the shape of open desk sessions of one hour organized regularly. A user can book such a session in an online form, providing a description of the image analysis problem they want to discuss, and uploading a few example images. During the preparation phase, the requests are each allocated to an open desk session, with a panel of 2 to 3 analysts selected according to their expertise. All the sessions happen the same day, with 2 to 3 running in parallel. The analysts join the user in an online meeting of 1 hour and work on the user’s problem during this time, actively prototyping the solutions they propose. The output of a session can be advice on how to address the problem, small analysis pipeline or small macros. \textbf{B.}~When a question cannot be addressed in one hour and / or requires some development, a user can request a collaborative project, during an open desk session or by direct contact. Project requests are then evaluated by the F-BIAS team, which estimates the project cost, start delay, and duration. Upon acceptance, an analyst is assigned based on availability and expertise. As a potential opportunity. a senior analyst from another site will provide mentoring to the project's analyst. The user and analyst collaborate asynchronously, with regular analyst reports, using various communication media (emails, online/onsite meetings). Project outputs may include analysis pipelines, results, and reports. }
    \label{fig:FigureN2}
\end{figure}

Users can request a collaborative project by direct contact, via Euro-BioImaging user access website or during an open desk session. 
The project registration form developed by Euro-BioImaging provides a template for users to specify the project content.
However due to the diversity of the projects handled by F-BIAS (which can, for instance, extend beyond light microscopy), it is challenging to create a universal template that perfectly captures all requests while remaining concise. 
As a result, we always discuss with the user before making any commitment. 
The request is then discussed among the F-BIAS team - that serves as a selection committee - to evaluate the project’s feasibility. 
If deemed feasible, one analyst with the required expertise and availability takes over the project. 
It evolves iteratively with regular discussions between the biologist and the analyst. 
Additionally, the project’s analyst may be paired with another analyst from a different site providing technical mentorship. 
This again results in capitalizing on user services to improve the skills and expertise of analysts (Figure~\ref{fig:FigureN2}B). 
Given the current overbooked status of analysts, projects are allocated solely based on their expertise and availability, and potential outcomes have only little impact in the decision process.

Unlike consultations, collaborative projects are billed. 
We use an hourly fee to charge the user for the projected effort of the project. 
When the actual effort exceeds the planned duration by a significant amount, a new quote is prepared in agreement with the user.
We leveraged our virtual and decentralized organization to establish a simplified economic model for F-BIAS. 
In the inception phase described here, we rely on the existing recharge systems implemented by contributing core facilities.
The quote for a project handled by an F-BIAS analyst, funded centrally by FBI, is sent from the local core facility in which they are embedded, and the fees are received by the core facility. 
In turn, the core facility hosts and manages the analyst, handling all administrative overhead.
The hourly fees are subsidized by FBI to lower and standardize the project cost to the user, ensuring that all F-BIAS analysts bill the same amount regardless of the core facility they work in. 
This subsidy can take the form of funds or in-kind contributions, with the F-BIAS analyst funded by FBI dedicating part of their time to work on user projects for the local core.
This organizational structure is straightforward and has the advantage of requiring minimal effort to set up. 
This demonstrates the commitment and importance of a national infrastructure like FBI, which can leverage funds to support the dissemination of bioimage analysis services at the national scale.

\definecolor{LunarGreen}{rgb}{0.215,0.254,0.223}
\definecolor{Black}{rgb}{0,0,0}
\definecolor{Nandor}{rgb}{0.286,0.337,0.298}
\definecolor{Tundora}{rgb}{0.262,0.262,0.262}
\definecolor{AthensGray}{rgb}{0.972,0.976,0.98}
\begin{table*}
    \centering
    \footnotesize
    \begin{tblr}{
      width = \linewidth,
      colspec = {Q[252]Q[521]Q[25]Q[25]},
      cells = {m, fg=Tundora},
      row{odd} = {AthensGray},
      row{1} = {Nandor,fg=white},
      column{1} = {l},
      column{2} = {l},
      column{3} = {r},
      column{4} = {r},
      hlines,
      vlines = {Black},
      vline{1} = {-}{LunarGreen},
    }
    
    Project title 
    & Notes 
    & Effort (h)
    & Cnslt. (h) \\
    
    Characterization of motile bacteria trajectories from microscopy time-lapse images                                      
    & The project involved adapting a Python library developed by a F-BIAS member to compute trajectory descriptors to characterize a new bacteria species, and deploying it to the user's lab. The analyst also trained the user to perform tracking of the bacteria.            
    & 50                               
    & 5                       \\
    
    {Feasibility study on multi animal tracking tools on non-tagged individuals}                                          
    & Testing the feasibility of using selected tracking tools to track single, non-tagged fishes, in aquaria containing a dozen of them. The study results are used in a grant application to implement and refine these tools for fish behavioral studies.                
    & 60                               
    & 8                       \\
    
    Automatic measurements of frequency and amplitude of beating cilia in bronchial cells from high-speed microscopy movies 
    & The project was deemed infeasible due to the complexity of the image content. We tested several approaches quickly, that made us confident we would not address the request. We guided users to pursue collaboration with research groups focused on image processing. 
    & 4                                
    & 2                       \\
    
    Automated recognition and classification of red blood cells               
    & The goal is to build an end-user tool that can detect several classes of red blood cells in a large collection of images. The first part of the project involves identifying the optimal algorithm for red blood cell detection and classification.                    
    & 50                               
    & 3                       \\
    
    Custom development around an existing codebase 
    & Users required some changes to be made in the core of a scientific tool developed by a member of F-BIAS, so that they could build their own plugin based on it.     
    & 80                               
    & 10                      
    \end{tblr}
    \caption{Examples of several F-BIAS collaborative projects.  \textit{Effort}: Number of hours billed or worked on in total for the project. \textit{Cnslt.}: Number of hours of consultations associated with the project.}
    \label{table-projects}
\end{table*}

\section*{The added value of a distributed image analysis core facility}
\label{the-added-value-of-a-distributed-image-analysis-core-facility}

\subsection*{From the analysts' viewpoint}
\label{from-the-analysts-viewpoint}

F-BIAS successfully addresses isolation issues mentioned in the above. 
Within F-BIAS, analysts exchange knowledge and good practices, teach each other analysis techniques and share information on key publications, online resources or conferences. 
Moreover, effective communication about the ongoing projects helps prevent duplication of work on similar project, saving both time and resources.
Experienced members can also mentor early-career peers and help them navigate the global bioimage analysis world. 
Conversely, early-career analysts challenge the techniques senior analysts are used to relying on, and help them renew their technical portfolio. 
This is particularly valuable given the young age of the field in academia. 
The consultations with several analysts amounted to an opportunity for learning how others approach a new problem and exchanging ideas. 
It also provided the group’s developers with a chance to demonstrate new tools and developments, have them tested by other analysts, and receive feedback. 
F-BIAS is highly valued by its members as a vital platform for honing skills, exploring new techniques, and discovering new tools.

\subsection*{From users' viewpoint}
\label{from-f-bias-users-viewpoint}

The scientific community of France is the main target of this endeavor, especially biologists who do not have access to bioimage analysis expertise onsite, the most common situation in the country.
These scientists now have access to the expertise of about 20 analysts and the means to leverage it for their projects. 
Consultations are tailored for efficiency and maximizing the scientific output of a one-hour session. 
F-BIAS analysts master a wide range of image analysis languages and packages from scripting to advanced analysis modules and AI libraries. 
By describing their needs in the online registration form, users will be put in contact with analysts that have the given expertise to answer their question, when possible. 
From the user viewpoint, this process is much simpler than having to identify the required expertise to find the right facility to contact. 
This one-hour consulting service with several experts is free of charge and requires little effort beside registration and sending test data beforehand. 
Finally, when a question cannot be solved in one hour and the user does not have resources to develop a suggested solution, F-BIAS can initiate a collaborative project with the user. 
F-BIAS truly succeeds in overcoming the technological barriers associated with bioimage analysis for researchers within France.

\subsection*{From local users' viewpoint}
\label{from-local-users-viewpoint}

As mentioned above, host cores enable their analysts to contribute to F-BIAS due to the scientific added value received in return. 
Indeed, the improved skills and expertise of F-BIAS members indirectly benefit their local users. 
The diversity of image analysts’ backgrounds, from applied mathematics, computer science, biology to physics offers another opportunity as they now have indirect access to a much wider range of approaches in which they can tap in to tackle their image analysis problems. 
Additionally, when an analyst is faced with a complex problem from a local user, they can reach out to their peers in F-BIAS for assistance or advice.

\subsection*{For the France-BioImaging infrastructure}
\label{for-the-france-bioimaging-infrastructure}

The integration of F-BIAS within France-BioImaging enables it to offer an efficient service in image analysis that complements other services. 
Moreover, F-BIAS reached critical size to foster beneficial interactions with other projects of France-BioImaging related to bioimage analysis. 
For instance, the FBI.Data project is focused on scientific image data management, offering tools for intelligent data archival, high performance computing, and implementation of FAIR data principles.
A federated data management solution~\cite{FBIData-repo} is currently being deployed in the different nodes of the FBI infrastructure, in strong connection with shared computing resources.
FBI.Data can capitalize on F-BIAS to test and  deploy its solutions in host core facilities, and reach out to end users, using F-BIAS members as a medium. 
FBI also supports the development of large scientific framework for image analysis such as Icy~\cite{de_chaumont_icy_2012}, BioImageIT~\cite{prigent_bioimageit_2022} and 
MorphoNet~\cite{leggio_morphonet_2019}. 
F-BIAS members serve as early adopters, beta testers and feedback source for these projects, offering a wide range of biological applications to disseminate the software through the F-BIAS user base. 
Finally, F-BIAS can redirect projects that fall outside the scope of its services to research groups of FBI focused on image processing.

\subsection*{For industry partners}
\label{for-the-industry-partner}

F-BIAS is in the best position to recommend an expert in image analysis to industrial companies, who can address their challenges. 
For now, this type of request is handled locally by the analyst's core facility in collaboration with the company.

\section*{Challenges in delivering bioimage analysis services remotely with a distributed team}
\label{challenges-in-delivering-bioimage-analysis}

\subsection*{Availability of human resources}
\label{availability-of-human-resources}

The most significant challenge is to \textit{secure human resources}, in particular for the collaborative project services. 
A major issue in operating the group and bioimage analysis core facilities in general is the limited number of dedicated analysts compared to the needs of the biological community. 
Indeed, the funding from the institutions to secure analyst positions dedicated to service are too restricted to have a sufficient workforce.
F-BIAS is a virtual structure, initially built on existing resources that are federated from host core facilities. 
At its creation, all members were already saturated or close to being saturated with requests from their local scientific community. 
This low availability is the main factor delaying image analysis projects~\cite{lewitter_need_2009}. 
The inception of F-BIAS was facilitated by the scientific value it offered to members and the minimal time commitment required for participation in consultations and exchanges. 
This helped in building and maintaining a sufficiently large team of willing analysts. 
Additionally, the positions granted by FBI for collaborative projects further strengthened our services.

The challenge with human resources is likely to endure. 
Ideally, an economical model is built around these services, and the fees billed for collaborative projects would help fund new positions offering these services nationally. 
However, the fees required for a full cost recovery would make F-BIAS services prohibitive for most research groups, and our main goal, democratizing bioimage analysis, would not be achieved. 
The availability of national funding for positions dedicated to the national infrastructure is instrumental in enabling the provision of services with a desirable capacity.

\subsection*{Diversity of host facilities goals and rules }
\label{diversity-of-host-facilities-goals-and-rules}

F-BIAS federates analysts from several core facilities, each with its own rules regarding core facility engineers job description, career, projects scope, projects billing, etc. 
In addition to the geographical spread, employers and institutions of these analysts are also different.
However, the same cost of services is offered to F-BIAS users regardless of which members contribute, so that the only factor determining what analyst is paired to a project is their skills.
To achieve this, we could again leverage on the FBI infrastructure, that subsidized F-BIAS prices, and the support of our host facilities that agreed on a homogeneous price.

\subsection*{Ensuring the scientific added value of internal meetings}
\label{ensuring-the-scientific-added-value-of-internal-meetings}

The scientific added value analysts find in F-BIAS is crucial for the success of the initiative. 
It helped convince the analysts’ managers to allocate time to the initiative, and helped retain analysts and prevent dropouts. 
It is therefore important to control the content of F-BIAS regular and annual meetings. 
We do this by enforcing the rules described above and focus on technical exchanges. 
Regular meetings are mainly focused on scientific content, asking for each other's help or showing a new technique, and round tables on more organizational aspects are reserved for our annual hackathons.

\subsection*{\texorpdfstring{Remotely operating}{Remotely operating}}
\label{remotely-operating}

A second challenge roots in the remote communication modality we rely on to deliver services. 
There are numerous factors that can degrade the quality of an online meeting, and they tend to occur frequently, such as low sound quality from bad equipment and unstable internet connection~\cite{fauville_zoom_2021}. 
These mere annoyances become prominent in the context of consultations, where we enforce the duration of a session to be one hour. 
To mitigate them the users are requested to upload the images and to check their communication equipment in advance. 
We observed that sessions tend to be inefficient when they do not.

\section*{The desirable components of a distributed image analysis core facility}
\label{the-desirable-components-a-distributed-image-analysis-core-facility}

There are a few other initiatives building nationwide services for bioimage analysis. 
Among them we can mention the \textit{SciLifeLab BioImage Informatics infrastructure} in Sweden~\cite{noauthor_scilifelab_nodate} and the \textit{National Facility for Data Handling \& Analysis in Italy}~\cite{noauthor_national_nodate}. 
They have the advantage of enjoying human resources dedicated to these projects from their inception. 
SciLifeLab has units at various sites (not all analysts are based at the same university), with agreements in place to govern personnel responsibilities, payments, and other related matters.
On our side we started as a virtual structure, built from existing resources initially with other duties, spread over the country. 
Our approach described above managed to circumvent these challenges and even converted them into opportunities. 
Nonetheless we identified a few key components, additional to the human resources and online tools discussed above, that would facilitate reproducing such a service elsewhere. 
As we did not benefit from all of them immediately, we can attest to their importance.

\subsection*{Central budget}
\label{central-budget}

Having a unique budget on which to bill the projects and use it to organize workshops, training, buy shared computing resources or recruit engineers would allow for a smooth operation of the facility. 
It should also be possible to apply to grant calls as a unified team, without having to split the budget between several institutes, which would greatly simplify the process. 
Working under the umbrella of a nationwide infrastructure such as FBI was very beneficial in our case, as it could address these needs. 

\subsection*{Project selection committee and selection criteria}
\label{project-selection-committee-and-selection-criteria}

We have begun advertising our services cautiously, and it is likely that the current number of project requests discussed above does not yet fully represent the actual demand in the territory we aim to cover. 
As our resources are insufficient for user requests, we envision that we will have to prioritize requests in the near future. 
Consequently, the criteria for accepting projects and establishing their priority order must be determined, and a selection committee composed of team members must be established.

\subsection*{Open desk policy and organization tools}
\label{open-desk-policy-and-organization-tools}

The organization of remote open desks has been more challenging than expected. 
To not tax too much of the engineers time and still provide a quality service to the life scientists, the session schedule has to be optimized. 
Users not showing up, or data not transferred beforehand are the most common issues that waste time. 
The transfer of data before the session should be enforced for maximum efficiency of the remote session. 
The registration form could be more comprehensive to really prepare the session beforehand. 
Such a form could be built by taking inspiration from the questions that should be asked for a bioimage analysis project~\cite{cimini_twenty_2023} but should take into account the fact that the user filling it might be quite inexperienced with quantitative approaches and may be unfamiliar with the vocabulary. 
Specializing a chatbot~\cite{lei_bioimageio_2024} to help the user filling the online form could be a solution to propose an interactive and user friendly registration and could even help to answer basic questions beforehand.

\subsection*{Internal data and code sharing tools}
\label{internal-data-and-code-sharing-tools}

As we grew, the need to share more than expertise arose. 
With the advent of deep learning, the availability of annotated training data has become a key factor for bioimage analysis. 
A distributed facility can capitalize on the diversity of dataset the analysts have access to, through the host institutes and the different equipment, samples and topics they have and focus on. 
In addition to sharing data, a distributed core facility will need to share code with its members, such as broadly applicable routines. 
The bioimage analysis community is used to distribute data and code publicly, but a distributed core facility offers the opportunity to share them early and privately with its members, while the code is being elaborated and the data is being annotated.
A few users may require not to make anything related to their project public before it is published.
A central resource for hosting code and data is therefore desirable for a distributed core. 
Currently, we use the public github repositories of the tools we work with.
For collaborative work, we primarily rely on the resources of FBI, particularly those from the FBI.Data project, and RENATER~\cite{renater} collaborative services (Table~\ref{table-tools}).

\definecolor{LunarGreen}{rgb}{0.215,0.254,0.223}
\definecolor{Black}{rgb}{0,0,0}
\definecolor{Nandor}{rgb}{0.286,0.337,0.298}
\definecolor{Tundora}{rgb}{0.262,0.262,0.262}
\definecolor{AthensGray}{rgb}{0.972,0.976,0.98}
\begin{table}
    \centering
    \footnotesize
    \begin{tblr}{
      width = \linewidth,
      colspec = {Q[252]Q[521]Q[25]Q[25]},
      cells = {m, fg=Tundora},
      row{odd} = {AthensGray},
      row{1} = {Nandor,fg=white},
      column{1} = {l},
      column{2} = {l},
      hlines,
      vlines = {Black},
      vline{1} = {-}{LunarGreen},
    }
    
    Functionality
    & Tool deployed \\
    
    Online chat service for instant communication
    & A Mattermost instance, deployed in the Nantes FBI \\
    
    Online meetings with F-BIAS members 
    & MS Teams private meeting with all members invited \\

    User registration page for open desk session 
    & An interactive page on FBI website \\

    Data transfer from user for an open desk session 
    & The FileSender tool of RENATER \\

    Online F-BIAS open desk session
    & Two online meeting rooms, using Rendez-Vous by RENATER \\

    Documents keeping tracks of open desk session content 
    & Shared Google spreadsheet and documents \\
    
    \end{tblr}
    \caption{The common tools currently used in F-BIAS for collaborative work and delivering services.}
    \label{table-tools}
\end{table}

\subsection*{Access to computing resources}
\label{access-to-computing-resources}

Some projects have large requirements for computing power, not always present in all of the core host institutes. 
The distributed facility can be a means to share the resources of one host institute with a member of the core, or to invest in them, within the scope of a consortium agreement encompassing the distributed core activities.

\section*{Conclusion}
\label{conclusion}

F-BIAS started with the aim of federating analysts scattered across the territory and is now a working, distributed national core facility for Bioimage Analysis, that offers value to its users and members. 
Since its creation, F-BIAS has offered 35 consultations. 
With the new resources, we can work on about 5 to 10 collaborative projects per year. 
The feedback we received from users through the consultations has been very positive and the collaborative projects we are working on are key to the research we contribute to. 
We are convinced that the success of this initiative is not particular to France and can be easily replicated in other conditions. 
Beyond the desirable key components listed in the previous section, we find the most important factor in the design of such an endeavor is to build and maintain a strong scientific added value for its members. 
High quality bioimage analysis services fundamentally depend on analysts and their skills. It is therefore crucial to create a structure that can attract and retain analysts, and expand their knowledge and skills. 
Our commitment to maintaining scientific value for the analysts of F-BIAS prevents its erosion, sustaining the analysts' interest in contributing to it since 2021. 
The services built upon F-BIAS benefit from the improving proficiency of the analysts, and in turn, the services are organized so that they foster the development of the analysts’ skills. 

\definecolor{LunarGreen}{rgb}{0.215,0.254,0.223}
\definecolor{Black}{rgb}{0,0,0}
\definecolor{Nandor}{rgb}{0.286,0.337,0.298}
\definecolor{Tundora}{rgb}{0.262,0.262,0.262}
\definecolor{AthensGray}{rgb}{0.972,0.976,0.98}
\begin{table}
    \centering
    \footnotesize
    \begin{tblr}{
      width = \linewidth,
      colspec = {Q[100]Q[100]Q[100]},
      cells = {m, fg=Tundora},
      row{odd} = {AthensGray},
      row{1} = {Nandor,fg=white},
      column{1} = {l},
      column{2} = {r},
      column{3} = {r},
      hlines,
      vlines = {Black},
      vline{1} = {-}{LunarGreen},
    }

    \textbf{Collaborative projects}
    & Local core facility
    & F-BIAS \\
    
    N. projects / year / analyst
    & 8.2 ± 2.6 (N=6)
    & 8.0 ± 0.0 (N=2) \\
    
    Project effort
    & 44.8 ± 43.3 hours (N=52)
    & 48.8 ± 27.9 hours (N=5) \\

    Project duration 
    & 33.9 ± 20.6 months (N=52) 
    & 7.3 ± 4.0 months (N=3) 
    
    \end{tblr}
    \caption{Statistics on collaborative projects comparing a local core facility (the Image Analysis Hub (IAH) of the Institut Pasteur) and F-BIAS.
    For the IAH data, we used the project management system deployed there (PPMS, Stratocore). 
    These statistics are built on data from collaborative projects for local users and excludes other activities (consultations, maintaining computers and commercial software, administrative work, development projects, and meetings).
    To calculate the number of projects per analyst per year, we counted the active projects in 2024, including both ongoing and completed projects. 
    We then determined the number of projects per analyst, weighted by the fraction of time each analyst dedicates to the IAH. 
    The project effort and duration were measured for projects completed after 2022, from the number of hours billed to users. 
    Duration was calculated as the time from project start to closure.
    For the F-BIAS column, we use the data from Table~\ref{table-projects}, using the billed time for the effort, and the actual duration of the 3 projects we completed.
    Two analysts dedicated 50\% of their time to collaborative projects in 2024, which gives a number of projects per analyst of 8 with the 8 projects we worked on this year.
    }
    \label{table-stats}
\end{table}

How do the services that F-BIAS offers compare to the ones a user can find in a local core facility, if they have access to one? 
Based on our experience we observed the following. Consultations are more frequent in local cores compared to those of F-BIAS. 
This allows local users to return multiple times to their local open desk sessions, refining the advice they receive as they implement it in their research. 
This effect of frequent open desks suits the iterative nature of imaging-based projects better, and could also be implemented within F-BIAS with additional open desk sessions, possibly with a rotation of the analysts. 
F-BIAS collaborative projects are similar in effort compared to local projects, and national and local analysts will work on roughly the same amount of projects per year (Table~\ref{table-stats}).
Local projects extend for much longer however, mainly because in the local core used in the comparison, projects stay open until they result in a publication when possible.
Also, F-BIAS projects tend to be shorter in duration because user requests are divided into several iterative projects, each concluding with a go/no-go decision.
For instance, ambitious questions start with a feasibility study that makes a first F-BIAS project, and is followed by another one for implementation if the first is successful. 
This reduces the risk for a user of paying a large amount of money upfront for an uncertain outcome. 

There are other kinds of services that can be found in local core facilities and that we do not offer yet. 
In addition to providing consultation or project service, a distributed facility is a resource for training both the analysts and the users. 
Most engineers are already giving regular individual or group training in their host institutes or via networks such as GloBIAS. 
Despite these efforts, there is still a huge demand for more training events from users~\cite{miura_survey_2021, sivagurunathan_bridging_nodate}. 
A structure like F-BIAS could be utilized to collect funds and organize a few days of intensive image analysis classes, targeting locations where such courses are rare.
Local core facilities also offer access to shared workstations, with commercial and open-source software for image analysis installed and maintained.
This facilitates autonomous image analysis by the users, and by pooling costly resources in the core facility, enables cost reduction for the host institute. 
A remote version of this service for open-source software would be highly valuable, considering the challenges of properly deploying these tools.
For this a remotely operating core could leverage cloud-based solutions, deployed on computing resources paid for with a central budget, and maintained by a member of the national core with dedicated time for this task. 

This prompts for a comparison between the national and local analyst roles.
The services we offer in F-BIAS are currently limited to consultations and collaborative projects, while when we act as analysts for our local cores, we are also tasked with delivering the supplemental services listed above. 
But the main differences arise from daily interaction opportunities.
A local analyst will have the chance to develop frequent formal and informal interactions with their users, in the scope of a collaborative project or in recurring consultations.
This will help the analyst build deeper professional relationships with the local scientific community they support, and develop knowledge in the biological domain of application of the projects.
This in turn can accelerate and enrich future collaborations with the same community. 
A national analyst engages with possibly many scientific communities (as can be inferred from Table~\ref{table-projects}).
While this is stimulating, it poses a challenge in developing deep expertise across all these domains.
Additionally, an analyst working remotely will interact with their users solely through the medium of the project and only for the duration of the project.
This makes it difficult to build a sense of togetherness and shared purpose with the overarching research endeavor of an institute.
These challenges are common to any team delivering remote services. 
In our current model, however, they have a minimal negative impact.
F-BIAS is a virtual, distributed infrastructure and its members are part of research institutes from which they operate and reach a wide audience.
As such we are also part of a local team that help mitigate the negative effects of remotely operating for analysts.
As stated above this model comes also with straightforward project management, which is a significant benefit. 
Building on these advantages, local core facilities will continue hosting analysts funded by FBI, contingent on future developments.
Another key connection for collaborative projects is with the local imaging and microscopy cores. 
Imaging-based projects are highly iterative, and the input of these experts is crucial for their success.
It is important therefore that a user brings both analysts and imaging specialists in the project discussions, particularly when these discussions involve experimental design.
Within the scope of F-BIAS, we have found no significant difference between local and remote projects on this topic. 
The efforts required to connect with imaging specialists or to have users involve them in discussions are similar for both local and remote projects.

F-BIAS will need to address several organizational points if its services are to be sustained over the long term.
For instance, who will become responsible for maintaining a tool or pipeline we delivered when an F-BIAS analyst leaves the group?
So far, this has not been required with the tools developed so far but this is an important aspect to consider for future projects.
As stated above, the allocation of a project to an analyst is made considering availability and expertise, but what if a project with highly promising data and collaboration potential arises, and multiple experts are interested?
These points overlap with those relevant to local core facilities. 
Since F-BIAS is a virtual infrastructure deeply rooted in these local cores, we will be able to draw inspiration from them to address these issues.

Bioimage analysis services are perfectly compatible with a remotely operating core.
They can be efficiently delivered thanks to modern IT tools, and the Covid crisis forced the integration of remote collaboration into our work culture. 
At the same time, they have a very strong value to research, as we could infer from the feedback of researchers that do not have access to them. 
Initiatives such as F-BIAS can easily leverage a few scattered resources to create significant value for the territory it supports simply by federating them. 
As a prospective catalyst for innovation and collaboration in bioimage analysis, we can envision several extra activities beyond services to increase our value to the scientific community. 
F-BIAS is well-positioned to serve as an expert panel for industry partners, particularly those connected to FBI. 
Additionally, such infrastructures are well positioned to become a hub for additional resources dedicated to supporting the research of a country. 
They can become the recipient of infrastructure grants or resources from national research agencies.
These resources will help bridge the gap in resources we observe with bioimage analysis while making them available nationwide.

\begin{acknowledgements}
\noindent F-BIAS is supported by the Agence Nationale de la Recherche through France-BioImaging (ANR-24-INBS-0005 FBI BIOGEN, core funding, funds for M. Anselmet and A. Meslin) and the institutes that host its members, listed in the authors affiliations. 
M. Breuilly is funded by LABEX Cortex (ANR-11-577 LABX-0042) of University Lyon 1, within the program “Investissements d’Avenir” (ANR-11-578 IDEX-0007).
We are particularly grateful to the executive team of FBI, notably Caroline Thiriet, Jean Salaméro, Édouard Bertrand, Alexandre Philips and René-Marc Mège for their support and impulse in creating the structure. 
We thank our respective managers and board of directors that backed the creation of F-BIAS and encouraged us to contribute to it. 
We thank Laurent Essioux for critical comments on this manuscript.
We thank Nicolas Goudin for his involvement in the structure.
The figures use icons that are derived from the work of Hilmy Abiyyu Asad (the Noun Project) and released under CC BY-3.0. 
\end{acknowledgements}

\begin{contributions}
\noindent G.L. wrote the first draft of the article. 
J-Y.T. created the figures, reviewed and edited the article, integrating inputs and contributions from G.L., S.R., V.B., M.Amb., T.P., J.M., P.P-G., A-S.M and M-S.P. 
All authors reviewed and validated the paper and contributed to the creation of the core and its activities.
\end{contributions}

\section*{References}
\bibliography{F-BIAS-Paper-bibtex}

\end{document}